\documentclass{PoS}
\usepackage{subcaption}
\usepackage{cleveref}
\usepackage{circuitikz}
\usepackage{xcolor}
\usepackage{physics}
\usepackage{gensymb}
\usepackage{amsmath}
\usepackage{lineno}
\usepackage{wrapfig}
\definecolor{leonard}{RGB}{0,0,0}
\definecolor{christian}{RGB}{0,0,0}
\definecolor{felix}{RGB}{0,0,0}

\title{The POCAM as self-calibrating light source for the IceCube Upgrade}

\ShortTitle{The POCAM as self-calibrating light source for the IceCube Upgrade}

\author{
The IceCube Collaboration\footnote{For collaboration list, see PoS(ICRC2019) 1177.}\\
{\itshape \href{http://icecube.wisc.edu/collaboration/authors/icrc19_icecube}{http://icecube.wisc.edu/collaboration/authors/icrc19\_icecube}}\\

E-mail: \email{cfruck@ph.tum.de, felix.henningsen@tum.de, christian.spannfellner@tum.de}}

\abstract{
The planned IceCube Upgrade, consisting of seven new instrumentation strings, will be installed at the South Pole within 2022/2023. The focus of this upgrade is calibration, reduction of systematic uncertainties and atmospheric neutrino physics. Within this scope, the "Precision Optical Calibration Module" (POCAM) will be installed at a number of positions on these new strings, to act as a calibration light source. The POCAM is an in-situ self-calibrating, isotropic, nanosecond light source that emits flashes of adjustable intensity and pulse duration. The isotropy is achieved using a teflon integrating sphere which further allows the calibration of the total number of emitted photons per pulse, using the integrated sensors. Prototypes have been deployed and operated within the GVD telescope in Lake Baikal and within the STRAW experiment in the Pacific Ocean. We present POCAM results and experiences from the GVD and STRAW installations as well as first IceCube sensitivity studies and the following design prospects for this next-generation POCAM iteration.\\

\vspace{4mm}
{\bfseries Corresponding authors:}
\speaker{C. Fruck$^{1}$},
F. Henningsen$^{1,2}$,
C. Spannfellner$^{1}$
\\
{$^{1}$ \itshape Physik-Department, Technische Universit\"at M\"unchen, D-85748 Garching, Germany}\\
{$^{2}$ \itshape Max-Planck-Insitut f\"ur Physik, D-80805 Munich, Germany}
}

\FullConference{36th International Cosmic Ray Conference -ICRC2019-\\
		July 24th - August 1st, 2019\\
		Madison, WI, U.S.A.}

\begin{document}

\section{The Precision Optical Calibration Module}
\label{sec:pocam}
The \textit{Precision Optical Calibration Module} (POCAM)~\cite{Boehmer:2018zml}\cite{Resconi:2017mad}\cite{Krings:2015aab} is a self-calibrating light-emitter device for the \textit{in-situ} calibration of large-volume neutrino detectors. Its primary goals are the reduction and better understanding of systematic experimental uncertainties within the detector -- most notably the properties of the optical detection medium as well as efficiency and angular acceptance of the photosensor instrumentation. 
The limited understanding of the South Pole ice in IceCube results in a systematic uncertainty of the optical detector properties of around 10\%~\cite{Aartsen:2013rt}. It is one of the leading systematics that affects a number of analyses~\cite{Resconi:2017mad}. By providing in-situ calibrated light pulses, the POCAM aims to reduce these uncertainties down to a few percent.

\begin{figure}
    \centering
    \begin{subfigure}[b]{.55\textwidth}
        \centering
        \includegraphics[width=\linewidth]{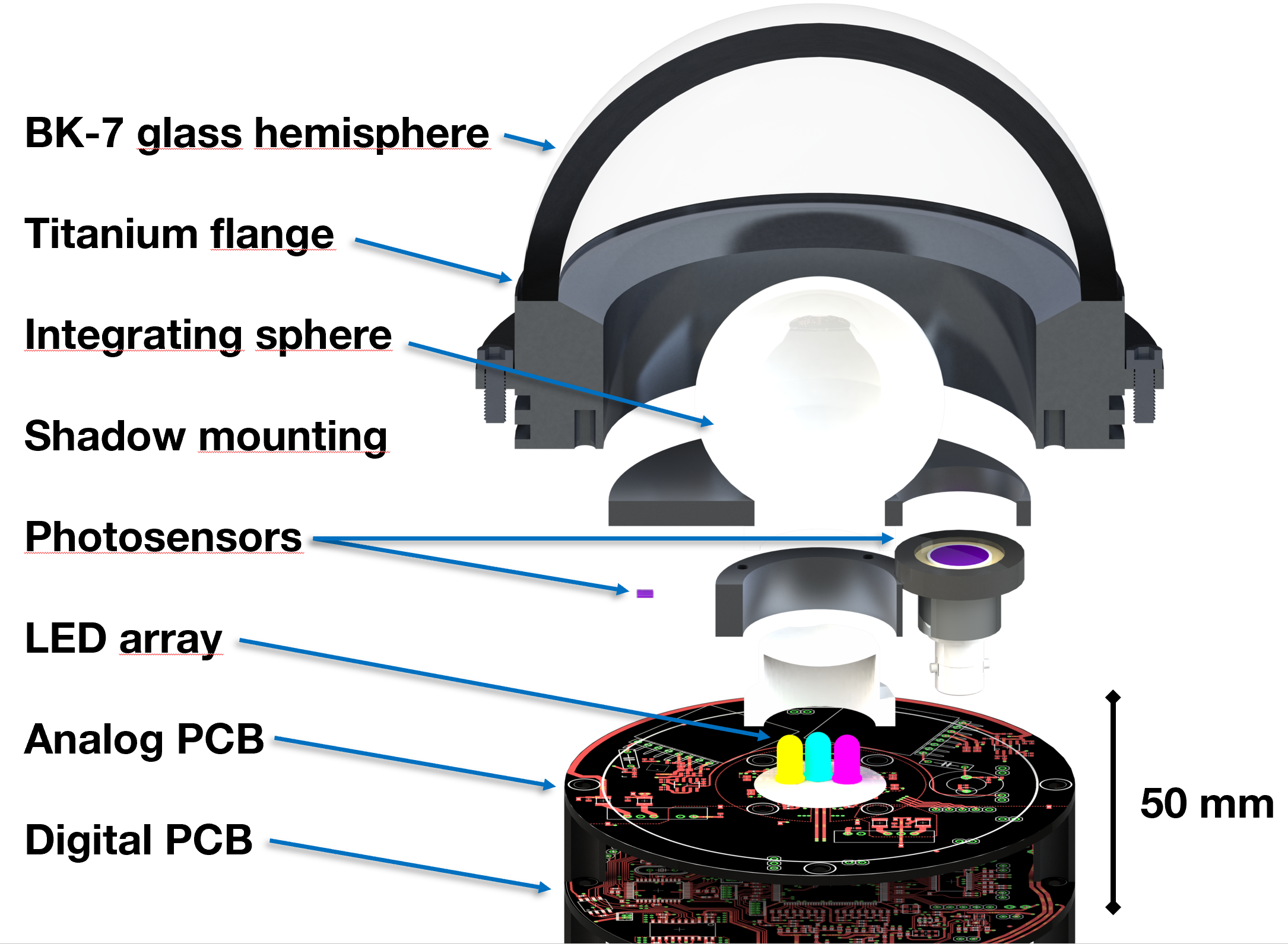}
        \label{fig:sub1}
    \end{subfigure}
    \hspace{0.1\textwidth}
    \begin{subfigure}[b]{.2\textwidth}
        \centering
        \includegraphics[width=\linewidth]{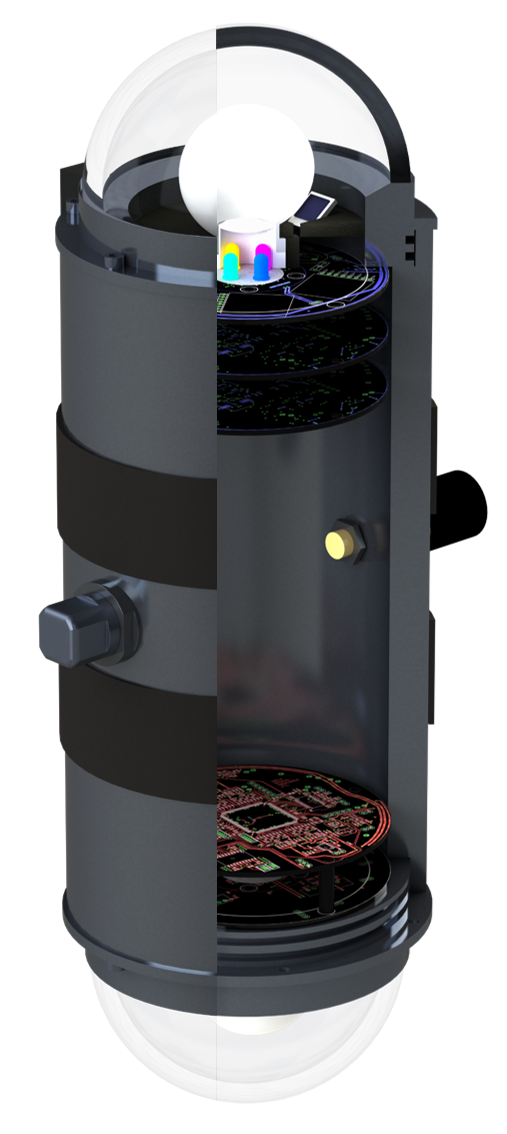}
        \label{fig:sub2}
    \end{subfigure}
    \caption{POCAM hemisphere assembly (left) and complete module assembly (right). The different components and their functions are explained in the text.}
    \label{fig:pocam-base}
\end{figure}

The baseline concept of the device is the isotropic multi-wavelength emission of nanosecond light pulses~\cite{Jurkovic:2016kxn} with the design shown in \cref{fig:pocam-base}. Here, the pressure housing from titanium with two glass hemispheres is shown together with internal components. The latter consist of a custom semi-transparent integrating sphere which makes the light pulse isotropic and the circuit boards for operation. The LEDs are predominantly driven by Kapustinsky drivers~\cite{KAPUSTINSKY1985612} and reach a dynamic range of $10^6 - 10^{9}\,$photons per pulse and pulse FWHMs of $1 - 8\,$ns, depending on the configuration. Usually LED wavelengths between $300 - 600\,$nm are used, with best intensity and timing performance achieved in the range of $400 - 470\,$nm. The integrated photosensors visible in \cref{fig:pocam-base} are a Silicon-Photomultiplier (SiPM) and a photodiode. These sensors monitor the pulse time onset and intensity characteristics in-situ with an anticipated precision of a few percent.
\par
As described in~\cite{Boehmer:2018zml} and~\cite{Resconi:2017mad}, the POCAM baseline design is composed of four sub-systems: pressure housing, analog and digital electronics and the light diffusion and emission components. The analog and digital boards are located in both hemispheres to allow for their independent operation. In the following we will describe the baseline design operating in the STRAW experiment~\cite{Boehmer:2018zml}. The device goals and design improvements for the IceCube Upgrade~\cite{ishihara2019:abc} will then be discussed in \cref{sec:icecube,sec:upgrade}. The upgrade itself will consist of seven new and more densely instrumented strings, deployed within a core segment of IceCube. They will include new photosensor instrument designs~\cite{classen2019:abc,nagai2019:abc} to increase sensitivity down to lower energies with more photo-active area as well as to deploy more instrumentation for better calibration, like the POCAM.

The pressure housing consists of a $10\,$mm thick and $25\,$cm long titanium cylinder and two titanium flanges to which the optical borosilicate glass is attached to using epoxy resin. The glass has a thickness of $7\,$mm and a diameter of $11.5\,$cm. The cylinder furthermore hosts the penetrator for electrical connection and a vacuum port for nitrogen-flushing of the instrument. The entire housing is manufactured by Nautilus Marine Service GmbH and is certified to a pressure resistance of $1500\,$bar, equivalent to a water depth of approximately $15\,$km. While this at first glance over-qualifies the housing for neutrino telescope applications in water, a deployment in the Antarctic ice of the IceCube detector can exert large pressures during refreezing of the instrument drill hole. As such, large safety factors are usually preferred in all marine operations and especially also for an application at the South Pole. In hindsight of that, all components have been chosen with temperature ratings down to $-55\,^\circ$C and are stress tested not only for temperature and pressure but also shocks and vibrations occurring during transport and deployment.

The analog boards in the POCAM electronics are responsible for the driving of light pulses and the readout of the integrated in-situ sensors. As such, it hosts the LED drivers, the photosensors, their frontend readout chain and related peripheral components. Additionally, the temperature of the LEDs, their drivers and the sensors is monitored, as this is part of the reference calibration described in \cref{sec:upgrade}. The main circuit used in previous POCAM versions (\cref{sec:deployment}) is the Kapustinsky driver~\cite{KAPUSTINSKY1985612}. It makes use of a controlled switching of two bipolar transistors which discharges a capacitor through an LED with a parallel inductance. The latter will cause a bipolar swing of the pulse and thus effectively cuts it short. With this technique and selected LED-types of $365$, $405$, $465$ and $605\,$nm, we achieved FWHMs of $4 - 8\,$ns with intensities ranging from $10^6 - 10^{9}\,$photons per pulse which is linear with supply voltage~\cite{Boehmer:2018zml,Lubsandorzhiev:2004zh}. However, the performance of the driver is highly dependent on the used type of LED and requires a proper selection process.
The photosensor frontend includes a charge-sensitive amplifier and an additional amplification stage for shaping of the output pulse. The output is a voltage pulse with few hundred nanosecond rising time and microseconds of decay time. This output is equal for both the SiPM and the photodiode but differs only in the sensitivity of the charge amplifier. Eventually the output signals are fed to the digital board.

Communication, data aquisition (DAQ) and control of the instrument are handled on the digital board which is also located in each of the hemispheres. Each digital board hosts primarily an FPGA, a micro-controller (MC) and a two-channel ADC (analog-to-digital converter). The ADC samples the output of the sensor frontend with a rate of $10\,$MHz and writes the data to storage, which in this case was an on-board SD-card. Additionally it provides orientation sensors based on accelero- and magnetometers as well as environmental monitoring with a number of temperature, pressure and humidity sensors overseeing the overall system integrity of the instrument. Communication with the instrument and data transfer was realised using a serial RS-485 interface and its operation is controlled via ASCII commands. The digital electronics and interfaces are by design modular and can be adapted to the respective application.

Finally, the primary and central part of the POCAM light emission is the diffuser. This semi-transparent integrating sphere from PTFE (or teflon) makes use of its Lambertian reflection to make the light pulses isotropic. It furthermore uses a dedicated two-part geometry of a plug and sphere that removes any disturbances on the emitting surface which would cause luminosity variations. The plug is further used to couple the LED array into the integrating sphere and uses a thin layer of teflon to diffuse the initial LED emission profile. This configuration can be seen in \cref{fig:pocam-base}. Thanks to the constant behaviour of teflon across a broad range of visible wavelengths, this allows to integrate visible light pulses from effectively $200 \,$nm upwards and effectively any LED opening angle. This emission behaviour has been extensively studied and the resulting zenithal profile is shown in \cref{fig:isotropy} for a diffuser made from different types of teflon. As can be seen there and also will be discussed in \cref{sec:upgrade}, switching from regular to specifically-produced optical teflon results in a significantly improved emission profile.

\section{Deployments in GVD and STRAW}
\label{sec:deployment}
A first prototype of the instrument has been deployed in 2017 within the GVD neutrino telescope in Lake Baikal~\cite{Resconi:2017mad}. The device was mounted to one of the detection strings and lowered to its final depth of around $1,000\,$m. During its one year of operation, several runs with POCAM have been carried out and allowed us to gain experience with the in-situ device operation. Furthermore, the shared data of the flasher runs -- provided by the GVD collaboration -- allowed an estimate of the attenuation length in Lake Baikal to $(21 \pm 4)\,$m at $470\,$nm. This agrees well with previous measurements~\cite{Belolaptikov:1997ry,baikal2} and eventually provided the first proof of concept for the POCAM.

The second deployment of an improved POCAM version was the STRAW experiment in the Northern Pacific Ocean~\cite{Boehmer:2018zml}. This pathfinder experiment deployed two $150\,$m mooring lines to a depth of $2.6\,$km in 2018 to probe the optical properties of the deep sea. These two strings host three POCAMs, the light emission of which is the baseline for the deduction of the optical parameters of the deep sea water using dedicated photosensors. Especially the modifications done with respect to the version deployed in GVD have improved the device operation significantly. Most notably, the light intensity was increased while the pulse width could be further reduced. This was achieved by performing characterization measurements on a large batch of LED types and manufacturers and selecting the best performing ones for the driver. Furthermore, the in-situ sensor front-end was optimized to reduce noise and the DAQ stability was improved. All details of these device characteristics can be found in~\cite{Boehmer:2018zml}.

\section{IceCube Upgrade Calibration Goals}
\label{sec:icecube}
The POCAM will be a multi-purpose calibration device as it can tackle both, systematics related to the South Pole ice as well as the surrounding sensor instrumentation. In total more than 20 of our devices will be distributed on the seven new Upgrade strings within the IceCube volume. These densely spaced and instrumented strings are in close proximity to previous IceCube strings and will host POCAMs spread between depths of $1450 - 2450\,$m. Further, for the first time, instrumentation, including POCAMs, will be installed in the deep ice down to $2600\,$m. This way, calibrations on old and new instrumentation can be performed within the regular IceCube volume as well as the vicinity volumes. As the POCAM is a multi-wavelength device, all systematic effects in the following, can be studied over a range of wavelengths from UV to visible.
\begin{wrapfigure}[19]{r}{9.75cm}
    \centering
    \includegraphics[width=\linewidth]{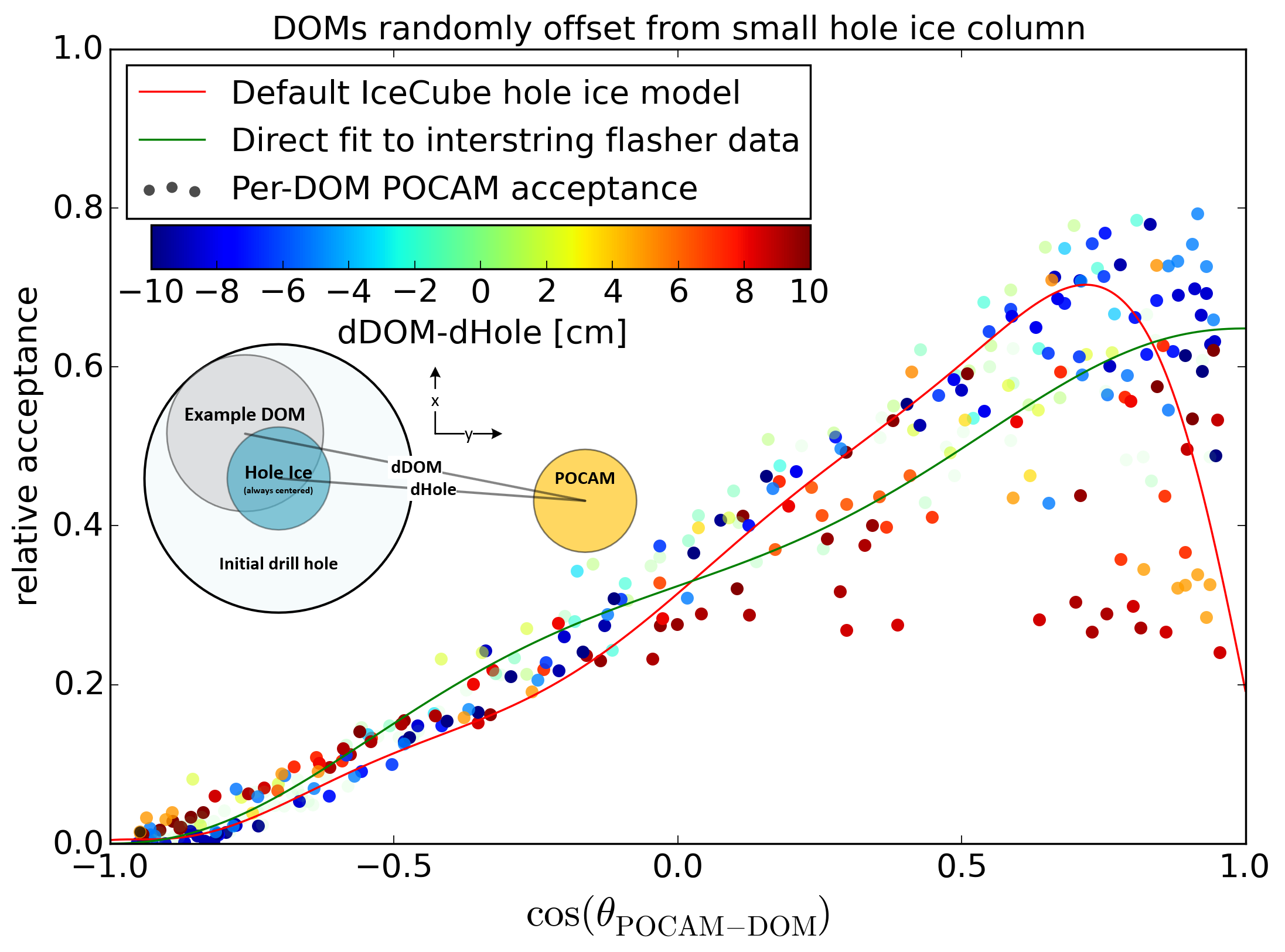}
    \caption{DOM angular acceptance curves for a strongly scattering hole ice column and DOMs randomly placed inside. Figure taken from~\cite{Resconi:2017mad}.}
    \label{fig:angularacc}
\end{wrapfigure}
One uncertainty in the understanding of the Antarctic ice of IceCube is the so called \textit{hole ice}. It describes a strongly scattering column of ice that forms during refreezing of the drill hole. This hole ice is a local ice formation that has a predominant effect on the angular acceptance profiles of the sensor instrumentation and is so far modelled empirically~\cite{Resconi:2017mad,Rongen:2016sbk}. Several POCAMs in the densely instrumented upgrade volume will be illuminating the same DOMs with effectively plane waves of light from different zenith angles. Thus, a sampling of the hole ice curve can be obtained on a per-DOM level, as is shown in \cref{fig:angularacc}.

\begin{figure}
    \centering
    \begin{subfigure}[b]{.45\textwidth}
        \centering
        \includegraphics[width=\linewidth]{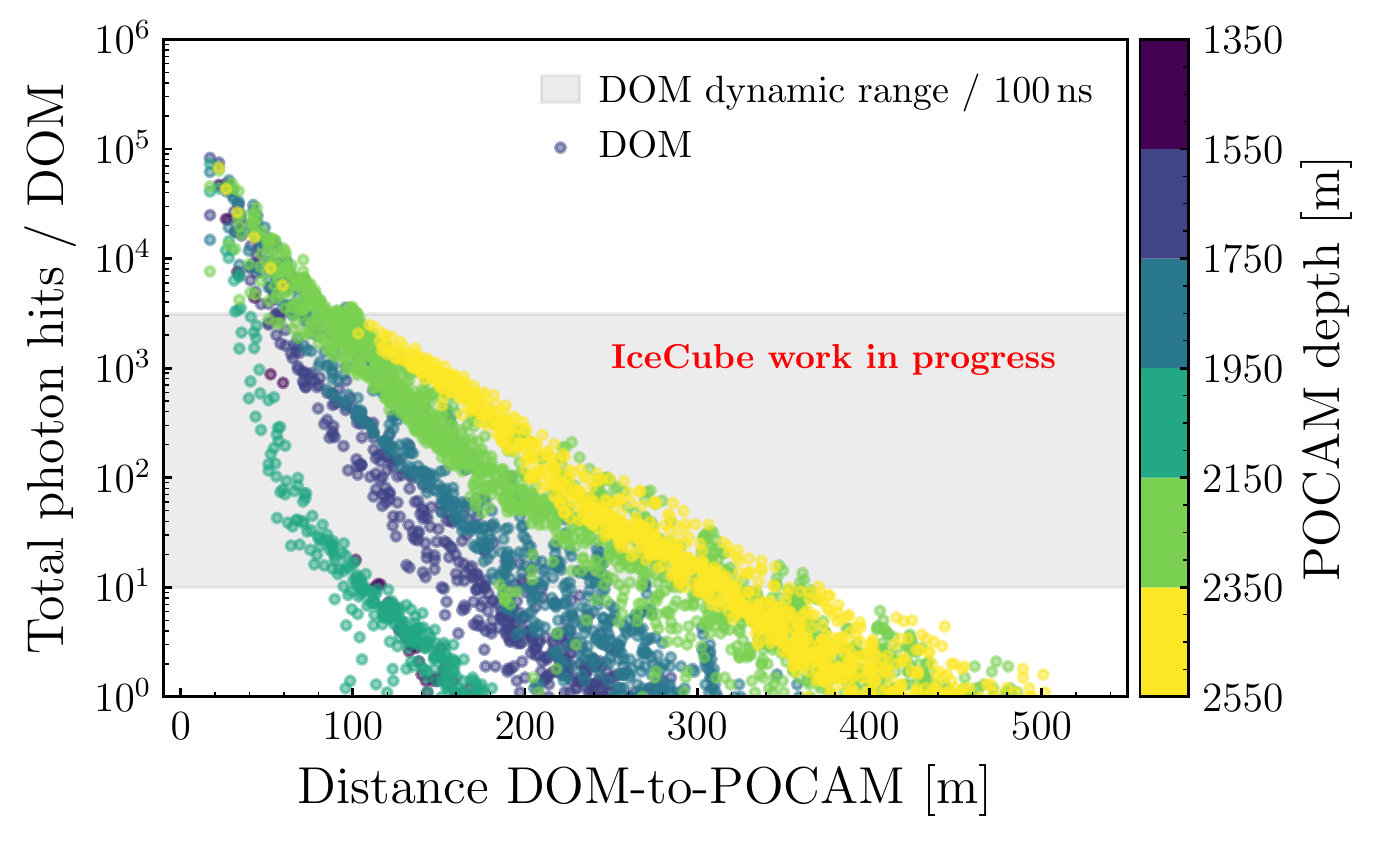}
        \caption{Average number of DOM hits by $N=1000$ isotropic $10^{10}\,$photon simulations at $400\,$nm and at different depths within the Upgrade volume.}
        \label{fig:dynamicrange}
    \end{subfigure}
    \hspace{0.05\textwidth}
    \begin{subfigure}[b]{.45\textwidth}
        \centering
        \includegraphics[width=\linewidth]{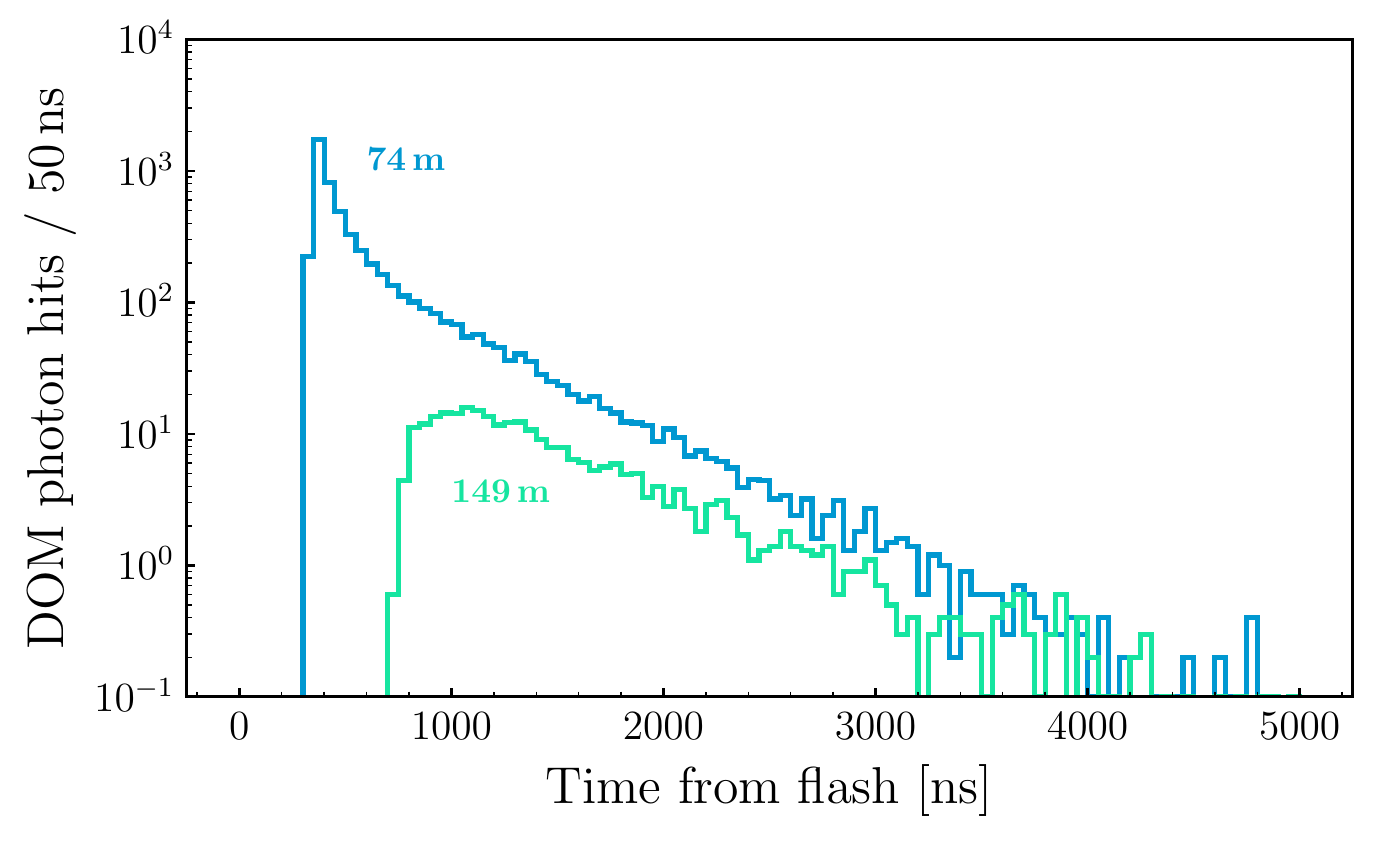}
        \caption{Photon arrival time histogram for DOMs at distances of $74\,$m and $149\,$m. }
        \label{fig:photonarrival}
    \end{subfigure}
    \caption{Driving effects of ice absorption and scattering on the POCAM light pulse properties. }
    \label{fig:baikal}
\end{figure}

Equally important for IceCube analyses are the bulk ice properties of the ice and the photodetection efficiencies of the DOMs. Both of these affect the global light propagation within the IceCube detector volume and its detection at the DOMs and thus are parameters of extensive studies~\cite{Aartsen:2013rt, Aartsen:2013ola}. The detector volume shows average scattering and absorption lengths of $20\,$m and $120\,$m respectively at $400\,$nm. As will also be discussed in \cref{sec:upgrade}, the POCAM will act as an absolutely calibrated light source with well-known emission characteristics. As such, the previously used LED calibration studies can be refined by using a well-defined light source in simulation studies that can be used to break the degeneracy of unknown light emission characteristics and ice properties. As far as the POCAM development is concerned, the dynamic range is the driving factor to illuminate a large enough volume with sufficient photon statistics. As shown in \cref{fig:dynamicrange}, this is achieved with flashes ranging from $10^5 - 10^{10}\,$photons per pulse. However, \cref{fig:photonarrival} shows that the strong ice scattering will diffuse out the timing of sharp light pulses, the time profile of the pulses can thus widen in time up to a few tens of nanoseconds in order to illuminate larger volumes.

A remaining open question is the ice anisotropy~\cite{Aartsen:2013ola,anisotropy2}. Its effect manifests itself as an observed anisotropic behaviour of the in-ice photon arrival time and flux which seemingly correlates with the direction of the ice flow. Using an isotropic light source and the surrounding DOMs, the POCAM will be able to study its influence by comparing azimuthal intensities, however, it requires precise knowledge of the orientation with respect to the hole ice column and the main cable.

\newpage
\section{Device Improvements for the IceCube Upgrade}
\label{sec:upgrade}

\begin{figure}[h!]
    \centering
    \begin{subfigure}[b]{.45\textwidth}
        \centering
        \includegraphics[width=\linewidth]{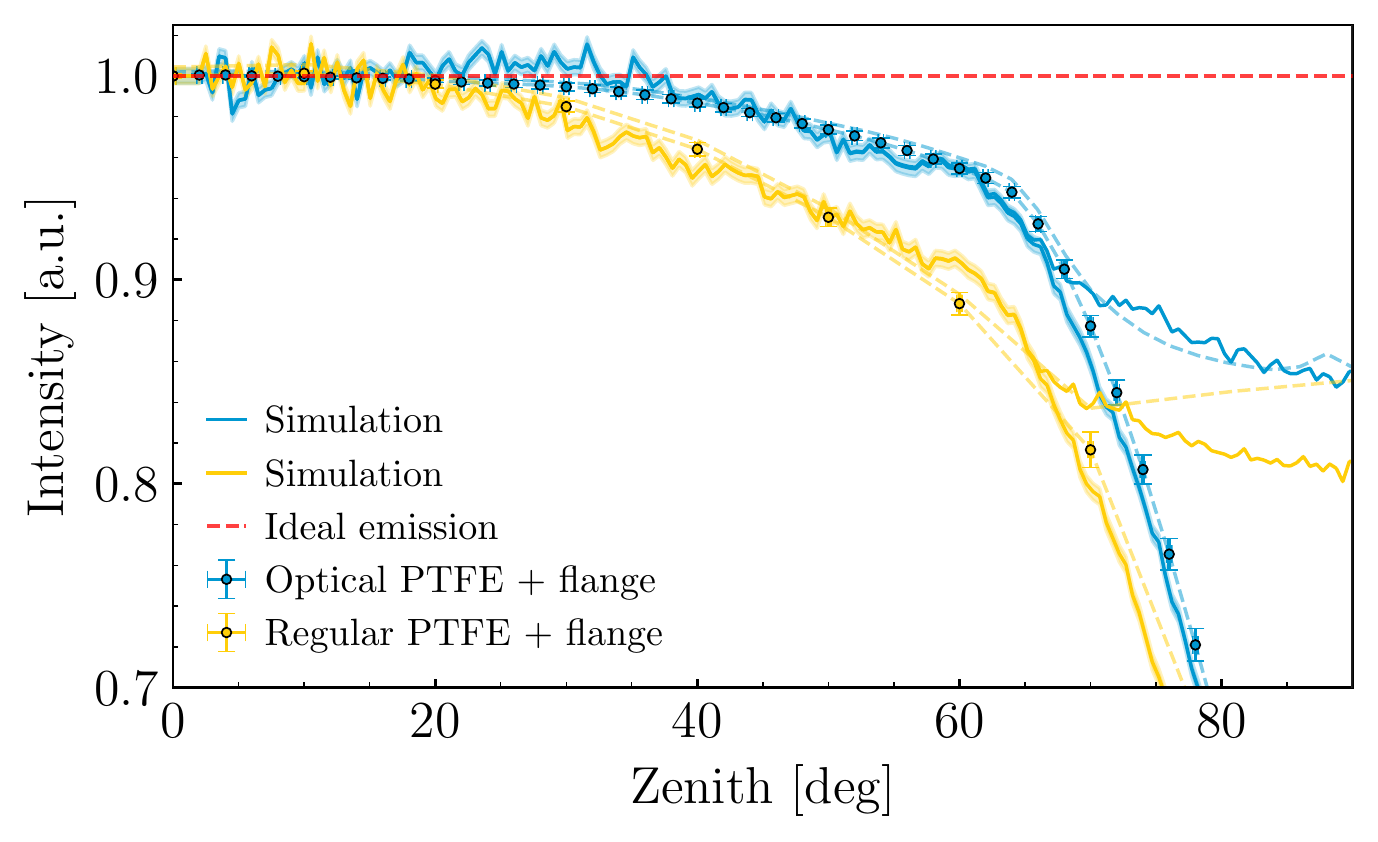}
        \caption{Zenith emission profile of a POCAM hemisphere using regular and optical teflon diffusers.}
        \label{fig:isotropy}
    \end{subfigure}
    \hspace{0.05\textwidth}
    \begin{subfigure}[b]{.45\textwidth}
        \centering
        \includegraphics[width=\linewidth]{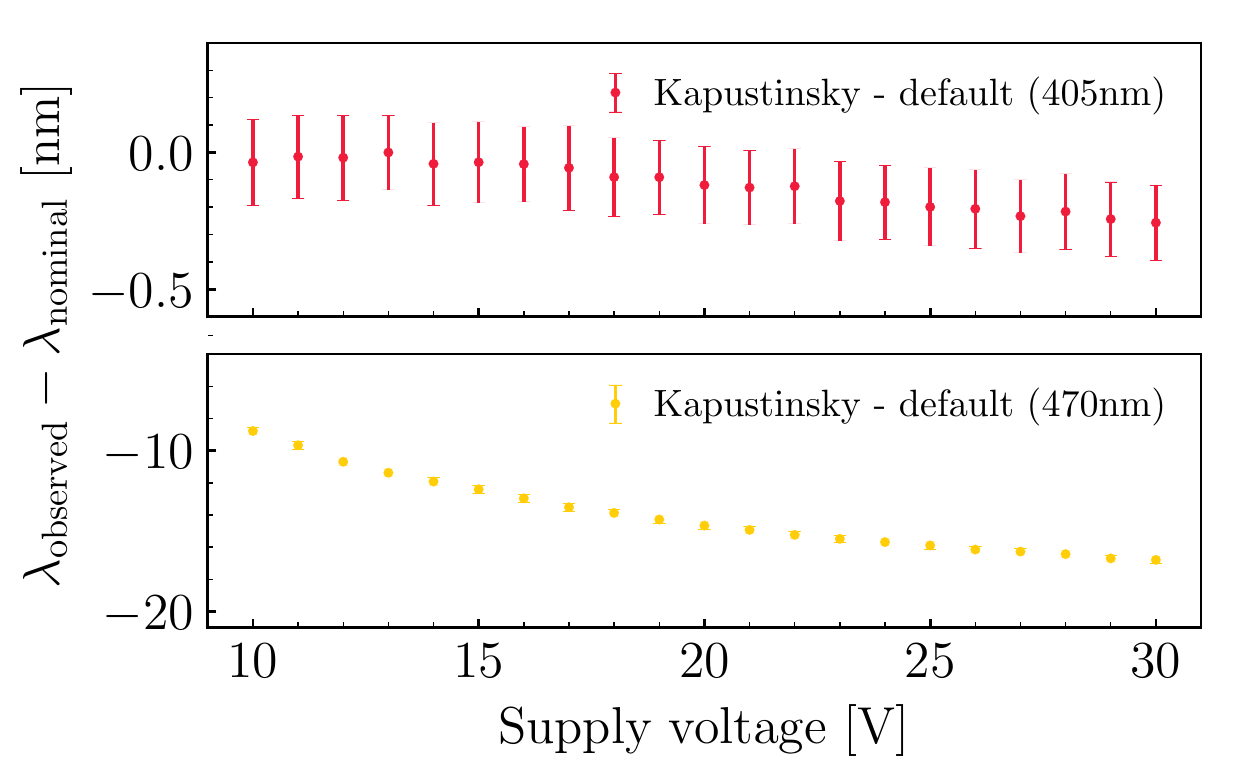}
        \caption{Mean emission wavelength versus voltage for a Kapustinsky driver.}
        \label{fig:spectrum}
    \end{subfigure}
    \caption{Measurements of integrating sphere emission profile and LED spectrum.}
    \label{fig:cal2}
\end{figure}

{\color{felix} While previous POCAM iterations have been succesfully deployed in environmental conditions similar to the South Pole, the instrument leaves room for precision optimizations necessary for an IceCube application. Components related to the light emission and in-situ calibration are hence subject to further improvements, this includes isotropy, dynamic range as well as in-situ sensor characterization and readout.

In order to illuminate large-enough fractions of the IceCube detector, the high-end of the dynamic range needs to be increased to $10^{10} - 10^{11}\,$photons per pulse. Therefore, new dedicated LED drivers are under investigation. They will operate alongside the fast Kapustinsky drivers and thus, are allowed to have pulse profiles up to a few tens of nanoseconds to provide necessary intensities. Promising systems -- based on the Gen1 driver of IceCube~\cite{Aartsen:2016nxy} -- using different gate drivers and GaN-FETs (Gallium-Nitride field effect transistors) are being developed. In addition, the default Kapustinsky driver will be accompanied by an additional one, tuned to sharp time profiles of the order of $1-2\,$ns FWHM but dimmer in intensity. Exemplary intensities and time profiles achieved with these baseline drivers are shown in \cref{fig:intensity,fig:timeprofile}, respectively.
\begin{figure}[h!]
    \centering
    \begin{subfigure}[b]{.45\textwidth}
        \centering
        \begin{tikzpicture}
   			\node[anchor=south west,inner sep=0] (image) at (0,0) {\includegraphics[width=0.99\textwidth]{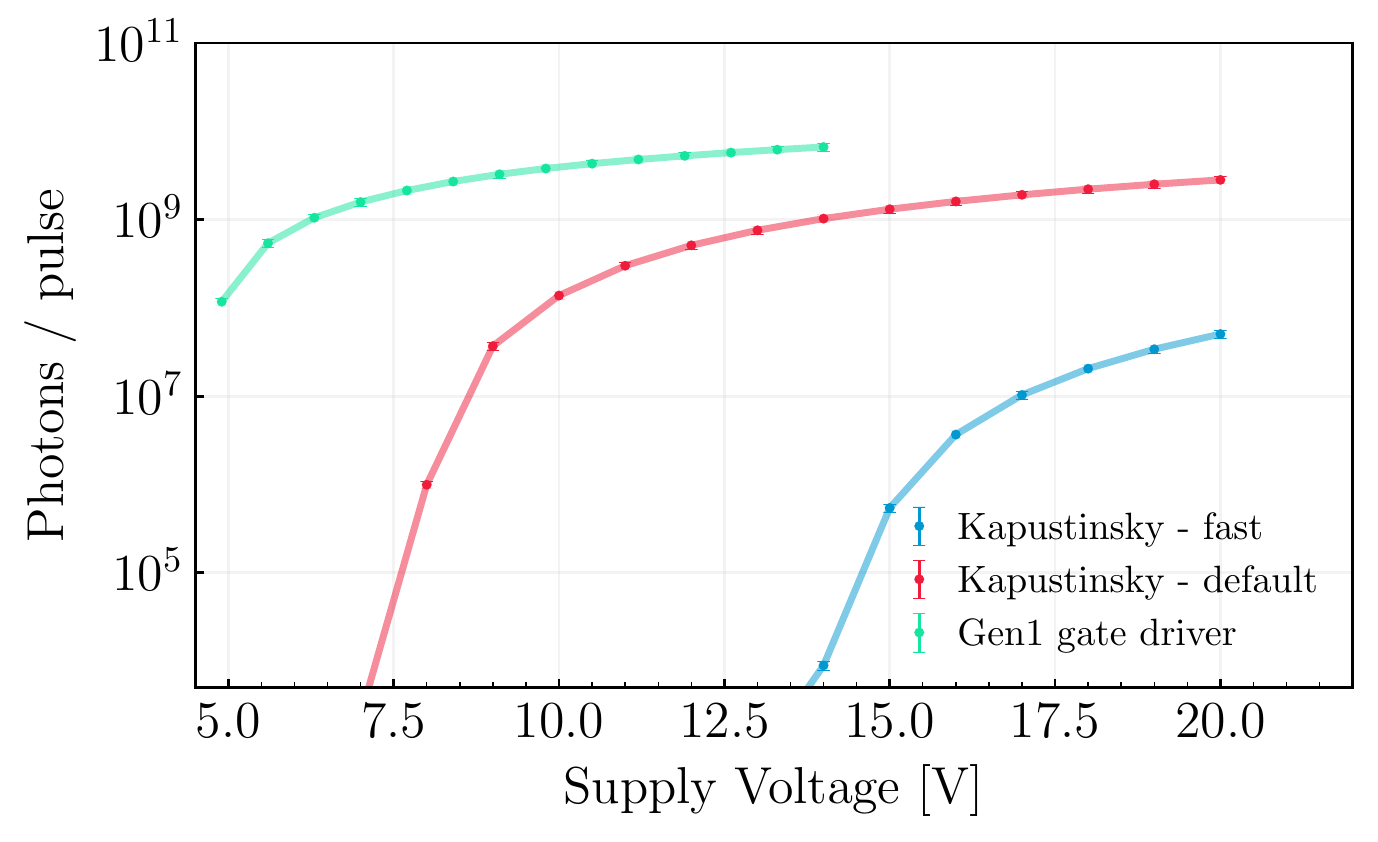}};
    		\begin{scope}[x={(image.south east)},y={(image.north west)}]
	        	\node at (0.825, 0.875)  {\tiny $\sim$10\% systematics};
	    	\end{scope}
		\end{tikzpicture}
        \caption{Intensity of POCAM LED drivers using the default $405\,$nm LED at different supply voltages.}
        \label{fig:intensity}
    \end{subfigure}
    \hspace{0.05\textwidth}
    \begin{subfigure}[b]{.45\textwidth}
        \centering
        \includegraphics[width=\linewidth]{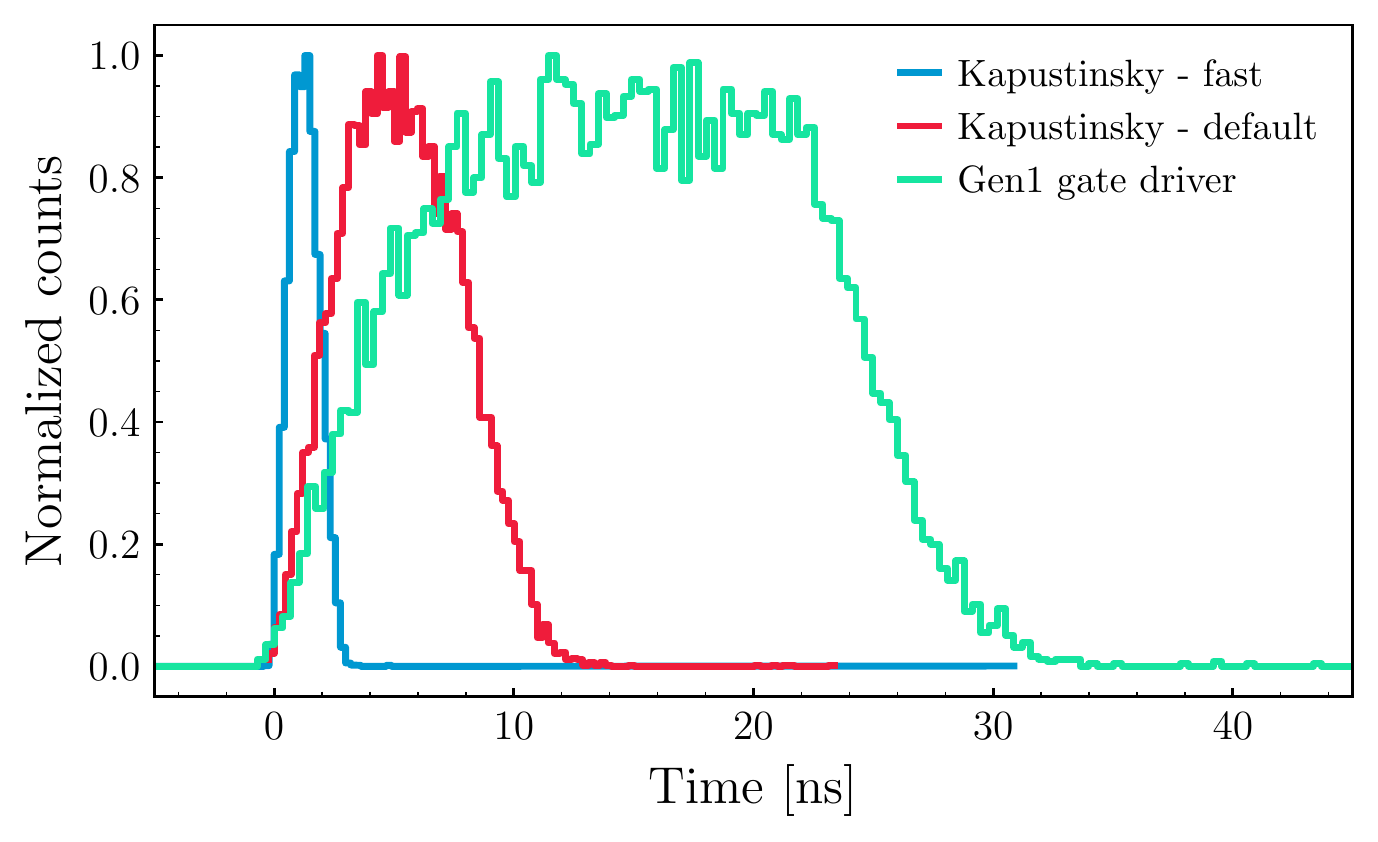}
        \caption{POCAM light pulse time profiles for different LED drivers and using the default $405\,$nm LED.}
        \label{fig:timeprofile}
    \end{subfigure}
    \caption{Measurements of light intensity and pulse shape for the default POCAM LED drivers.}
    \label{fig:cal1}
\end{figure}\par
In addition, the LED wavelengths will be adjusted for the IceCube Upgrade, with a default LED of $405\,$nm installed in every device. A secondary wavelength colour is selected from a spectral range of $320\,$nm to $550\,$nm and will depend on the final location of the POCAM in the ice, as the spectral interest in optical properties is depth-dependent. As mentioned in \cref{sec:deployment}, the right choice of the LEDs greatly affects the performance of the drivers. In the scope of finding the most suitable LED-type, automated characterization measurements on a large batch of LEDs will be performed to investigate their characteristics in connection with the driver. 
Also the readout electronics of the in-situ sensors are going to be revised with a new analog frontend. This will include new charge- and pre-amplifiers with significantly reduced bias current and thus higher signal-to-noise ratio for both SiPM and photodiode. In addition, the fast SiPM signal is split, a small fraction will trigger a giga-hertz comparator for optical onset pulse timing and the remainder is charge-integrated. Other components of the DAQ will stay as described in \cref{sec:pocam} with an additional board that will handle timing and communication with the IceCube detector and its protocols. As mentioned already in \cref{sec:pocam}, diffusers used in GVD and STRAW, were made out of regular teflon. As can be seen, in \cref{fig:isotropy} this achieves promising isotropic behaviour in zenith angle. In the scope of optimizations, a specifically produced teflon for optical applications was proposed. This optical teflon further increases the diffusion of the initial LED flash with only minor loss of intensity with respect to the regular one. However, \cref{fig:isotropy} shows that it notably increases the isotropy and as such is the material of choice for an IceCube application.}

In order for the POCAM to have the capability to calibrate itself, it is necessary to conduct precise characterization measurements of the light pulsers. These act as a reference during later operation. The main goal of this procedure is an automated relative characterization of the pulsers with respect to driver bias voltage and temperature. The latter poses more complicated, as it has to be carried out over a range of $-55\, \text{to} +25\,^\circ$C. Apart from an intensity and time profile calibration, it also becomes necessary to quantify the LED emission spectrum. As shown in \cref{fig:spectrum}, it has recently been observed that pulsed LED drivers vary in mean emission wavelength with applied bias voltage. This wavelength shift is important to account for different wavelength-dependent quantities within the IceCube detector, in order to achieve better calibration.
\par
{\color{leonard}
Together with the angular emission profile, four characterization measurements need to be conducted for every module. In order to streamline this process, two fully automated characterization stations are being developed. The first setup consists of a freezer, capable of reaching temperatures down to $-75\,\degree$C, the sensor instrumentation for the different characterization measurements and a reference light source for stability checks. The instrument electronics are mounted inside the freezer and the LED drivers are connected to the characterization sensors inside the dark box via optical fibers. The sensors include a spectrometer for the wavelength measurement, an avalanche photodiode to sample the time profile of the pulse via time-correlated single photon counting, as well as a photomultiplier for low and a photodiode for high intensities. Secondly, the angular emission scan is carried out with two rotational stages to which the POCAM hemisphere is mounted to. A photodiode measures the emitted intensity for different azimuth and zenith angles. All measurement points shown in  \cref{fig:cal1,fig:cal2} are carried out with prototypes of these setups.}

To conclude, the POCAM as an isotropic in-situ calibrated device will help to improve the understanding of the IceCube detector by providing means to reduce systematic uncertainties of the optical medium and the sensor instrumentation. The streamlined process of production and calibration will further ensure a precise knowledge of individual instrument characteristics.

\section*{Acknowledgements} 
We acknowledge the support by SFB1258 and the Cluster of Excellence. Additionally we thank the Baikal and STRAW collaborations for the deployments of the POCAM and sharing of related detector data. 

\bibliographystyle{ICRC}
\bibliography{references}
\end{document}